\documentclass[epj]{webofc}
\usepackage[varg]{txfonts}   % Web of Conferences font
\woctitle{Connecting The Dots / Intelligent Trackers 2017}
\usepackage{graphicx}
\usepackage[dvipsnames]{xcolor}
\definecolor{darkblue}{rgb}{0,0,.5}
\definecolor{darkgreen}{rgb}{0,.5,0}
\definecolor{darkred}{rgb}{.7,0,0}
\usepackage[thickspace, squaren]{SIunits}
\usepackage{amsmath}
\usepackage{commath}
\usepackage{mathtools}
\newcommand{\pT}{$p_\text{T}$}
\newcommand{\cphi}{$\varphi\,$}
\newcommand{\ctheta}{$\theta\,$}

\newcommand{\zaxis}{$z$-axis\,}

\newcommand{\xyplane}{$x$-$y$-plane }

\newcommand{\rzplane}{$r$-$z$-plane }
\newcommand{\eepair}{$e^+ e^-$}
\newcommand{\YfourS}{$\Upsilon\del{4S}$\,}

\begin{document}
\title{Online Data Reduction for the Belle II Experiment using DATCON}

\author{\firstname{Florian}   \lastname{Bernlochner} \inst{1} \and
        \firstname{Bruno}     \lastname{Deschamps}   \inst{1} \and
        \firstname{Jochen}    \lastname{Dingfelder}  \inst{1} \and
        \firstname{Carlos}    \lastname{Marinas}     \inst{1} \and
        \firstname{Christian} \lastname{Wessel}      \inst{1}\fnsep\thanks{\email{wessel@physik.uni-bonn.de}} 
}

\institute{Physikalisches Institut, University of Bonn, Germany          }

\abstract{%
The new Belle II experiment at the asymmetric \eepair accelerator SuperKEKB at KEK in Japan is designed to deliver a peak luminosity of $8\times10^{35}\text{cm}^{-2}\text{s}^{-1}$. 
To perform high-precision track reconstruction, e.g. for measurements of time-dependent CP-violating decays and secondary vertices, the Belle II detector is equipped with a highly segmented pixel detector (PXD). 
The high instantaneous luminosity and short bunch crossing times result in a large stream of data in the PXD, which needs to be significantly reduced for offline storage. 
The data reduction is performed using an FPGA-based Data Acquisition Tracking and Concentrator Online Node (DATCON), which uses information from the Belle II silicon strip vertex detector (SVD) surrounding the PXD to carry out online track reconstruction, extrapolation to the PXD, and Region of Interest (ROI) determination on the PXD. 
The data stream is reduced by a factor of ten with an ROI finding efficiency of >90\% for PXD hits inside the ROI down to 50\,MeV in \pT of the stable particles.

We will present the current status of the implementation of the track reconstruction using Hough transformations, and the results obtained for simulated \YfourS $\rightarrow \, B\bar{B}$ events.
}
\maketitle

\section{Motivation for the Belle II Experiment}
\label{sec:motivation}
Since the Belle experiment stopped taking data in 2010, its successor Belle II is being developed. 
The Belle II experiment will be located at the KEK laboratory in Japan, at the only interaction point of the SuperKEKB collider, which is the successor of the KEKB machine. 
SuperKEKB will collide asymmetric energy beams (7\,GeV electrons and 4\,GeV positrons) with a design instantaneous luminosity of $8\times10^{35}\text{cm}^{-2}\text{s}^{-1}$. 
This luminosity is 40 times the instantaneous luminosity of KEKB. 
This produces higher data rates and thus higher background rates with which the new Belle II detector has to cope. 
Due to limited bandwidth of the readout electronics and to minimise the amount of mass storage required, online data reduction is essential. 
The working principle of one of the two deployed online data reduction systems, the Data Acquisition Tracking and Concentrator Online Node (DATCON) is described in this article.

\section{The Belle II Vertex Detector}
\label{sec:belleIIdetector}
\begin{figure}[ht]
  \centering
  \begin{minipage}[t]{0.55\textwidth}
    \includegraphics[width=\textwidth]{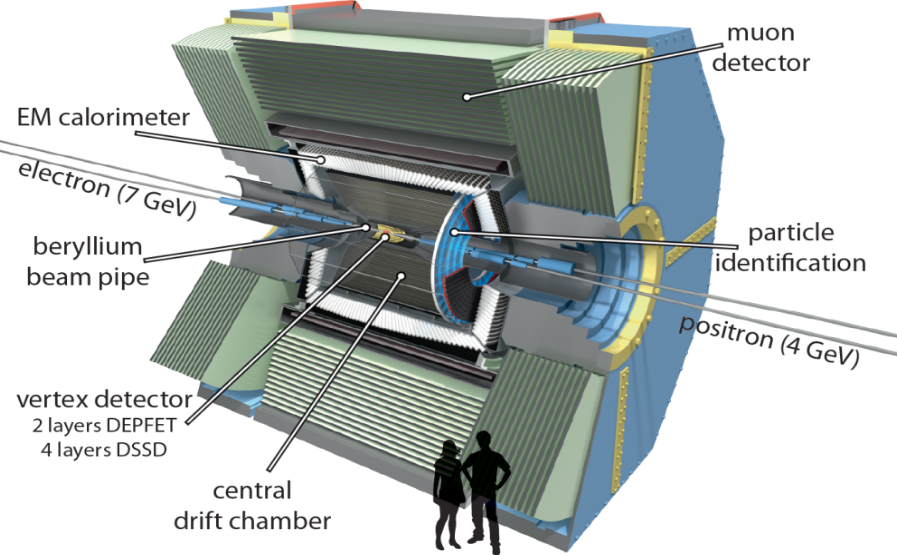}
    \caption{Overview of the Belle II Detector. }%\TODO{bibliography entry?} }
    \label{fig:BelleII}
  \end{minipage}
  \hspace*{\fill}
  \begin{minipage}[t]{0.4\textwidth}
    \includegraphics[width=\textwidth]{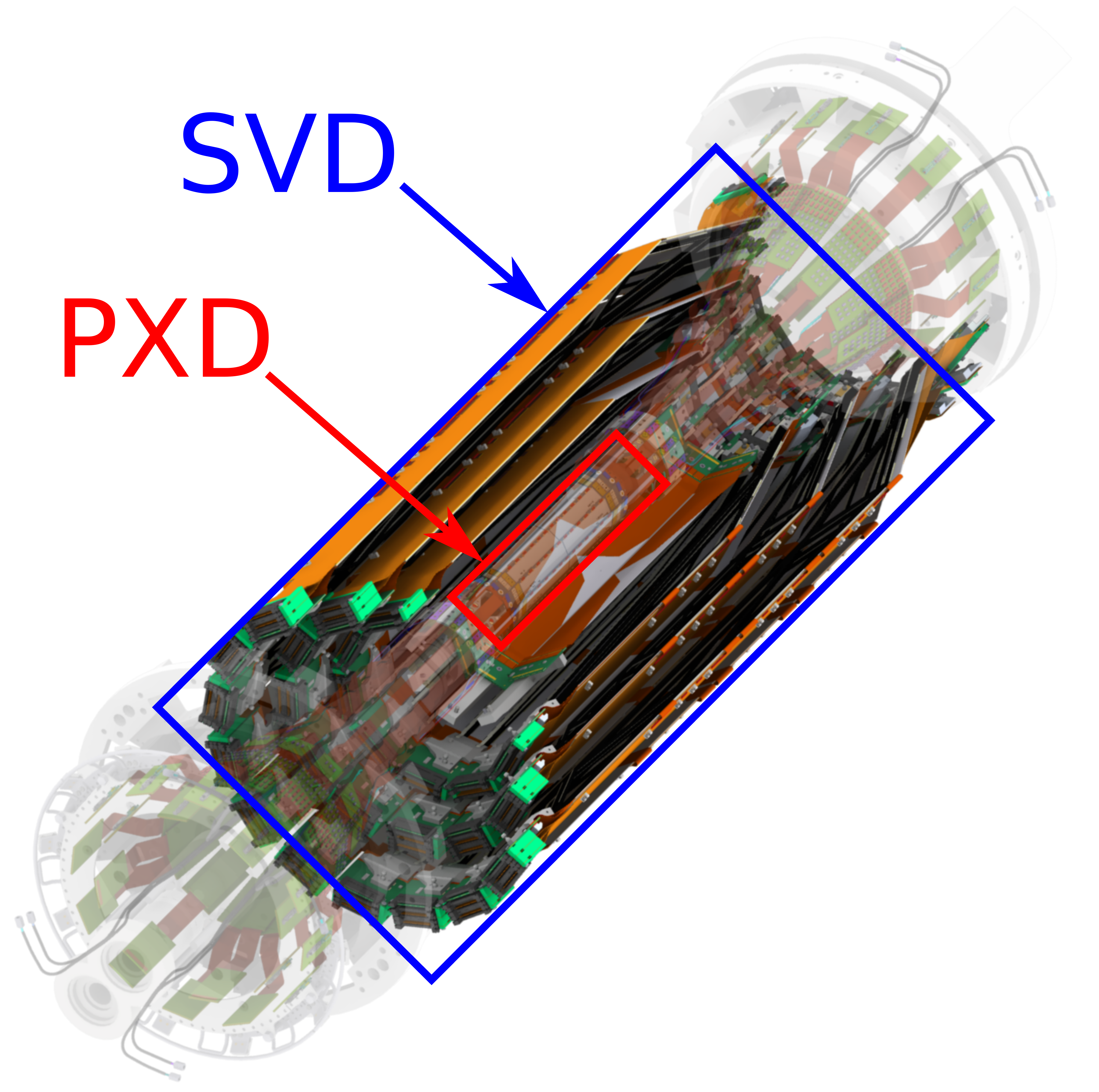}
    \caption{The Belle II Vertex Detector. 
      It contains two subdetectors, the PiXel Detector (PXD) with two layers, surrounded by the Silicon Vertex Detector (SVD) with four layers. 
      The innermost radius is 14 mm, the outermost radius is 135 mm.
    }%\TODO{bibliography entry?} }
    \label{fig:BelleIIVXD}
  \end{minipage}
\end{figure}

A sketch of the Belle II detector is shown in Figure \ref{fig:BelleII}. 
The innermost part of the detector is the silicon based vertex detector system (VXD), which is surrounding the beryllium beam pipe. 
The other sub-detectors are: the central drift chamber, the solenoid magnet producing a magnetic field of 1.5T, detectors for particle identification, the electromagnetic calorimeter, and an instrumented flux return that can detect $K_L$ and muons. 
The VXD consists of two components: a PiXel Detector (PXD) based on DEPFET \cite{DEPFET} technology and a Silicon Vertex Detector (SVD) based on double-sided silicon strip sensors. 
The placement of both components is shown in Figure \ref{fig:BelleIIVXD}.  
Both PXD and SVD are very thin detectors with only $0.2\%$ and $0.6\%$ of a radiation length per layer, respectively. 
The PXD consists of 2 layers at radii of 14 and 22\,mm from the interaction point with in total 40 modules. 
A module contains 250 $\times$ 768  pixels, having two types of pixels with a pitch of 50 $\times$ 55 $\mu \mathrm{m}^2$ (256 pixels in central region) and 50 $\times$ 70 $\mu \mathrm{m}^2$ (512 pixels in forward and backward region) for layer 1 and 50 $\times$ 65 $\mu \mathrm{m}^2$ (256 pixels in central region) and 50 $\times$ 70 $\mu \mathrm{m}^2$ (512 pixels in forward and backward region) for layer 2. 
The information in the 8 million pixels is read out at a data rate of 256\,Gb/s, corresponding to 90\% of the data rate of the complete Belle II detector without data reduction. 
The SVD consists of 4 layers at radii ranging between 39 and 135\,mm. 
In total the SVD contains 172 sensors with 768 $\times$ 768 strips having a pitch of 50 and 160 $\mu$m (layer 3) and 768 $\times$ 512 strips having a pitch of 75 $\times$ 240 $\mu$m (layers 4 to 6), respectively.
This results in about 240000 strips in total \cite{Abe:2010gxa}. \\

In a $B\bar{B}$ event coming from the decay of the \YfourS resonance there are on average 10 tracks in the acceptance region of the VXD, which covers the full angle of $2\pi$ in azimuthal direction (\cphi), and the region $17\degree < \theta  < 150\degree$ in polar angle. 
In addition to hits from particle tracks originating from $B\bar{B}$ events, background processes produce additional hits that are recorded by the VXD. 
Such background hits come for instance from two-photon QED processes, Touschek scattering, Coulomb scattering of beam particles with residual gas in the beam pipe, radiative Bhabha scattering and Synchrotron radiation. 
Beam-induced backgrounds increase the occupancy of the PXD up to 1\% and up to 0.25\% in the SVD, based on detailed simulations of the backgrounds. 
Only a small fraction of the hits in the VXD belong to tracks originating from the decay of $B\bar{B}$ pairs. 
The online data reduction of the Belle II experiment is designed in such a way that only the hits of interest for physics analysis recorded in the PXD are forwarded to permanent storage.
To implement this, the hits of other sub-detectors as the SVD are used to execute an online track finding. 
The reconstructed tracks are then used to define Regions of Interest (ROI) on the PXD and only the subset of pixels inside an ROI are permanently stored. 
For the maximum tolerable occupancy of 3\%, a data reduction factor of about 10 for PXD hit information is required. 

\section{Data Reduction Concept}
\label{sec:DataReduction}

\begin{figure}[!ht]
	\centering
  \includegraphics[width=0.8\textwidth]{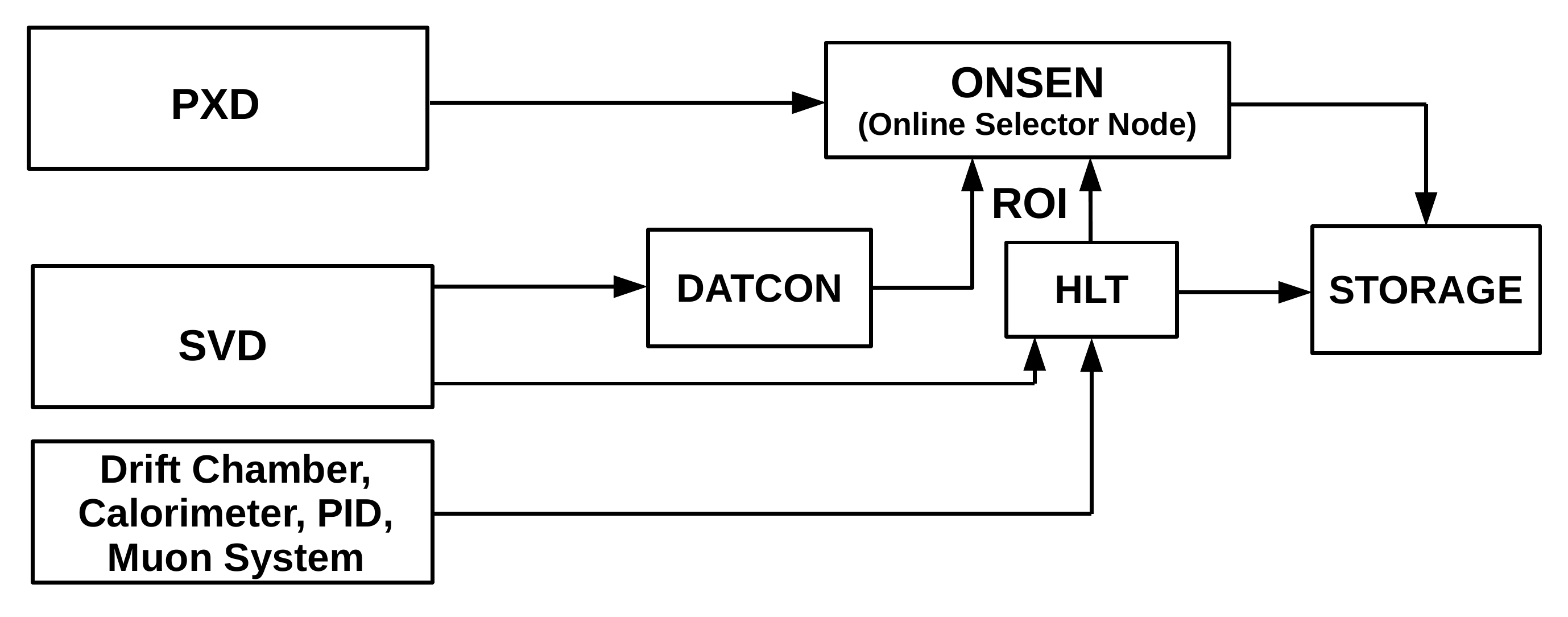}
  \caption{
  A simplified illustration of the Belle II data acquisition system with the two data reduction systems: 
  DATCON receives hit recorded by the SVD and performs an online track reconstruction to define ROIs on the PXD. 
  The HLT receives hit information from the SVD, the drift chamber, and the muon system as well as information from the PID detectors to define ROIs on the PXD. 
  Both HLT and DATCON run independently of each other. 
  The ROIs of both systems are sent to the ONSEN system, which performs the overall PXD data reduction by applying the ROIs to the PXD data. 
  }
  \label{fig:B2DAQ}
\end{figure}

Figure \ref{fig:B2DAQ} shows a simplified sketch of the Belle II Data Acquisition (DAQ) system. 
The data of the SVD, the drift chamber and the sub-detectors for particle identification (PID), calorimetry and muon detection are sent to the Higher Level Trigger (HLT) \cite{Itoh:2012via}. 
The data from the SVD are additionally sent to the DATCON, which performs online track reconstruction, extrapolation to the PXD, and calculation of ROI's in the PXD.
The ROIs are then sent to the Online Selector Node (ONSEN) \cite{Lautenbach:2017cnu}.
Besides DATCON, the HTL is the second system that performs a calculation of ROI's in the PXD. 
For this task the HLT not only can use data from the SVD, but also from the drift chamber and the other sub detectors. 
In addition, HLT provides the trigger signal for the complete detector and pipelines the data of the sub-detectors, except PXD, to the storage. 
The PXD data are only sent to ONSEN, which merges the ROI's of HLT and DATCON and performs the overall PXD data reduction by rejecting hits outside the ROIs.  
The HLT uses a computing farm with 6400 cores in total and runs sophisticated track finding and fitting algorithms. 
These HLT algorithms will also be used in the later offline track reconstruction. 
DATCON, on the other hand, runs a fast FPGA-based track reconstruction. 

\begin{figure}[!ht]
  \centering
  \includegraphics[width=0.7\textwidth]{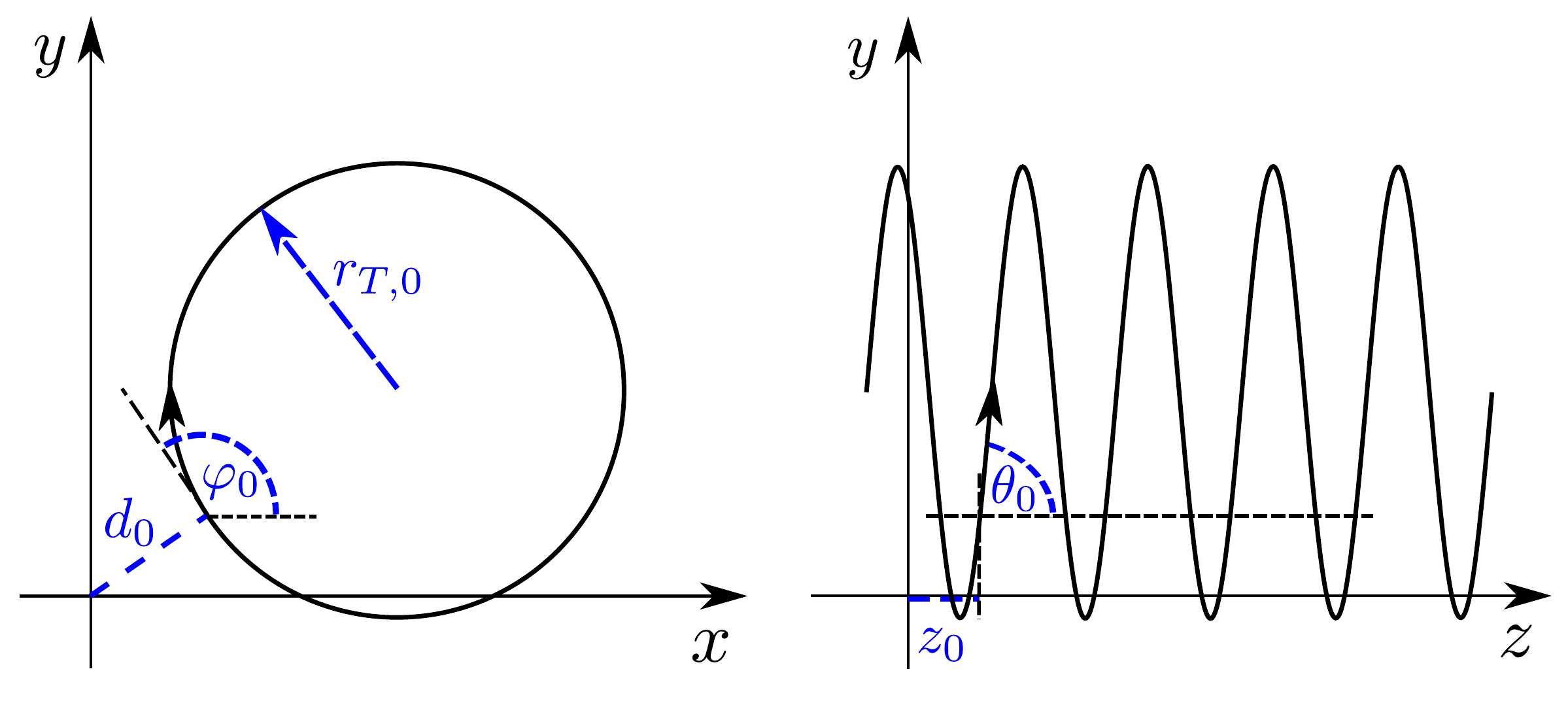}
  \caption{Two-dimensional view of a helix as parametrisation of a track inside a magnetic field. 
  The index 0 indicates parameters at the point of closest approach of an ideal track. 
  The values of these parameters can change with time due to e.g. multiple scattering  or other processes which cause energy loss. }
  \label{fig:helix}
\end{figure}
 
A track in a magnetic field, parametrised by a helix, is characterised by five parameters at the point of closest approach from to the beamline. 
This is illustrated in Figure \ref{fig:helix}. 
The parameters are: the initial azimuthal angle $\varphi_0$, the distance of closest approach $d_0$, the radius of the track $r_\text{T,0}$, the initial polar angle $\theta_0$, and the initial $z$-coordinate at the point of closest approach $z_0$.

\begin{figure}[!ht]
	\centering
  \includegraphics[width=0.6\textwidth]{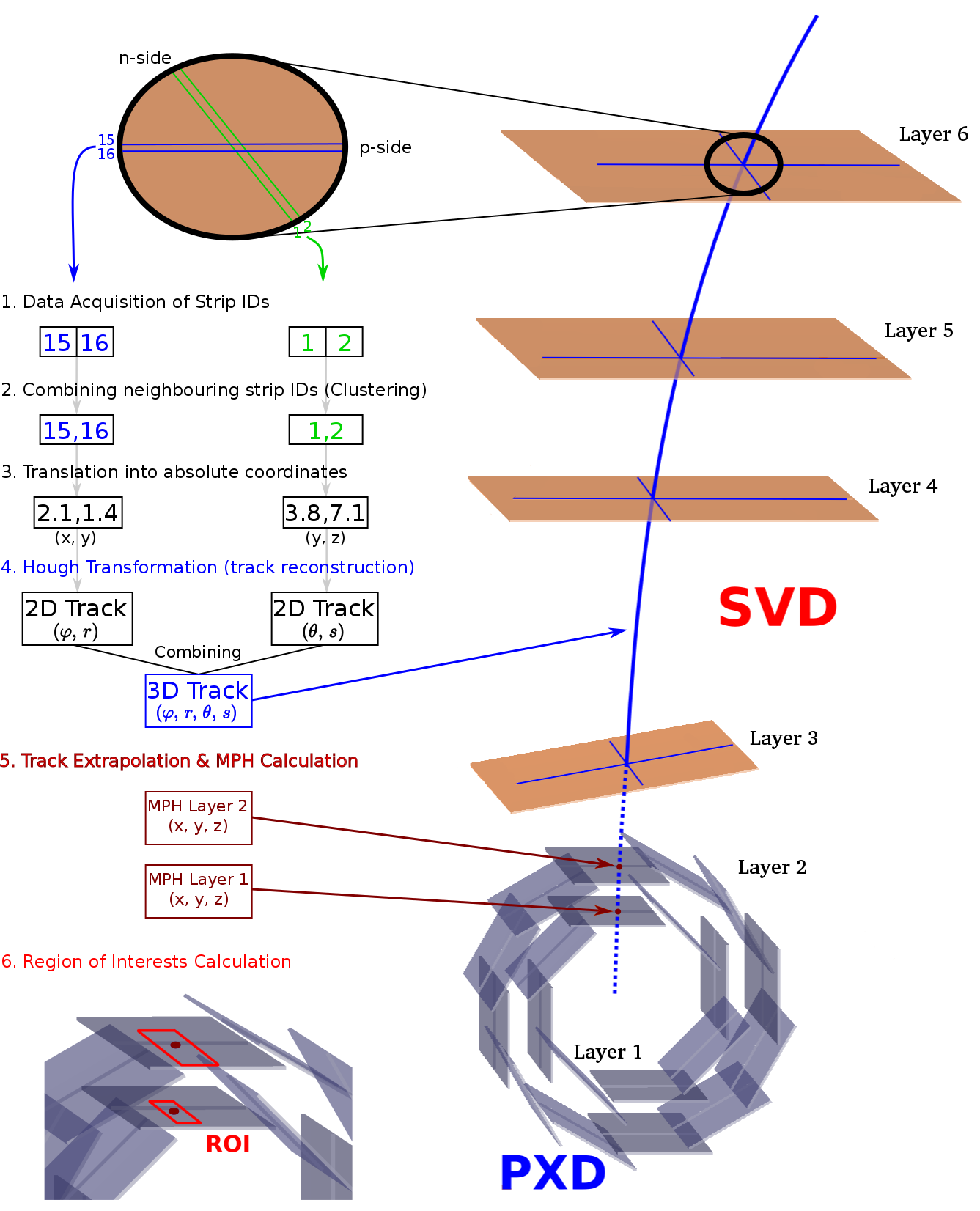}
  \caption{Schematic overview of the DATCON track reconstruction and extrapolation to the PXD. 
  The data of the the SVD strips (1.) are converted to clusters (2.) in local detector coordinates, from which the absolute coordinates of the hits are computed (3.). 
  These coordinates are Hough-transformed (4.) to reconstruct possible tracks, yielding two 2D tracks with information on \cphi and $r$, and \ctheta and $s$, respectively. 
  The information is combined to obtain 3D tracks, which are extrapolated to the PXD. 
  The positions of the most probable hits are calculated from the intersection of the extrapolated tracks with the PXD layers (5.). 
  Finally the ROI's on the two PXD layers are calculated (6.). 
  }
  \label{fig:reductionscheme}
\end{figure}

A schematic overview of the data reduction procedure and the data flow inside DATCON is shown in Figure \ref{fig:reductionscheme}: 
The ROIs are calculated by using SVD hits to reconstruct tracks and extrapolating the trajectories to the PXD.
For the purpose of track finding, the tracks are assumed to be circular in the \xyplane, with an additional ghost hit at the origin of the coordinate system where the beam spot is located.
The algorithm assumes that the trajectories can be approximated by straight lines in the \rzplane with $z_0 \neq 0$. 
The hits are transformed using Hough \cite{Hough:1959qva} and Hesse transformations:
\begin{align}
  \label{eq:rho}
  \rho \ &= \ \frac{2}{r_\text{hit}^2} \cdot \left( x' \, \cos \varphi \, + \, y' \, \sin \varphi \right), \\
  \label{eq:s}
  s \ &= \ r_\text{hit} \, \cos \theta \, + \, z_\text{hit} \sin \theta,
\end{align}

where $\rho \, = \, \frac{1}{r_\text{T}}$ denotes the track curvature, $r_\text{hit}$ is the distance of the hit from the \zaxis,
$z$ denotes the global $z$ coordinate of the hit, and $\left(x', \, y'\right)$ are the conformally transformed values of $\left(x, \, y\right)$ of the hit defined by:
\begin{equation}
	\label{eq:conformal}
  \left(x', \, y'\right) \ = \ \frac{\left(x_\text{hit}, \, y_\text{hit}\right)}{r_\text{hit}^2} \ = \ \frac{\left(x_\text{hit}, \, y_\text{hit}\right)}{x_\text{hit}^2 \, + \, y_\text{hit}^2}.
\end{equation}
The conformal transformation is only valid for $d_0 = 0$, which is a good approximation for $B$-meson decays as their decay products (except the charged particles $e^\pm$, $\mu^\pm$,  $\pi^\pm$, and $p/\bar{p}$) only have a short life-time and decay in close proximity to the origin of the \zaxis. 
Note that the conformal transformation is needed as the Hough transformation is better applicable to straight lines. 
Using conformal and Hough transformation, a helix trajectory is mapped onto a straight line. 
Hits on this straight line correspond to intersecting of lines in the Hough parameter space. 
Thus the task of finding tracks in real space is equivalent to finding intersections of lines in Hough space. 
Once the intersections are found, equations \ref{eq:rho} and \ref{eq:s} allow for a straightforward computation of the angle and track curvature $\rho \, = \, \frac{1}{r_\text{T}}$. 

With this information intersections of all tracks with the PXD detector planes are calculated. 
This reduces the task of finding ROIs to finding the intersection of a circle with a straight line in two dimensions in the \xyplane and to finding the intersection of two straight lines in the \rzplane. 
From these interceptions, Most Probable Hits (MPH) are defined and a fixed-size ROI of $80 \times 120$ pixels is created. 
Studies to tune the ROI size on the estimated particle momenta are currently being performed. 
Finally, the identified ROIs are transmitted to ONSEN. 

\section{Preliminary Results}
\label{sec:development}

To develop and test the necessary algorithm to be implemented on an FPGA, a C++ and python based implementation of the algorithm is used, running inside the Belle II Analysis and Software Framework (BASF2) \cite{Kim:2016mak}. 
The BASF2 framework allows for the simulation of the detector response of particles traversing the Belle II detector and includes the full decay chain of \YfourS\,$\rightarrow \, B\bar{B} \, \rightarrow$ stable particles (i.e. $e^\pm$, $\mu^\pm$,  $\pi^\pm$, $p/\bar{p}$, and $K_L$). 
This allows one to assess the performance of a given algorithm using simulated decays and variable background conditions.
The preliminary performance of such simulations of the DATCON using 100,000 simulated $B\bar{B}$ events are shown.

\begin{figure}[!ht]
  \centering
  \includegraphics[width=0.7\textwidth]{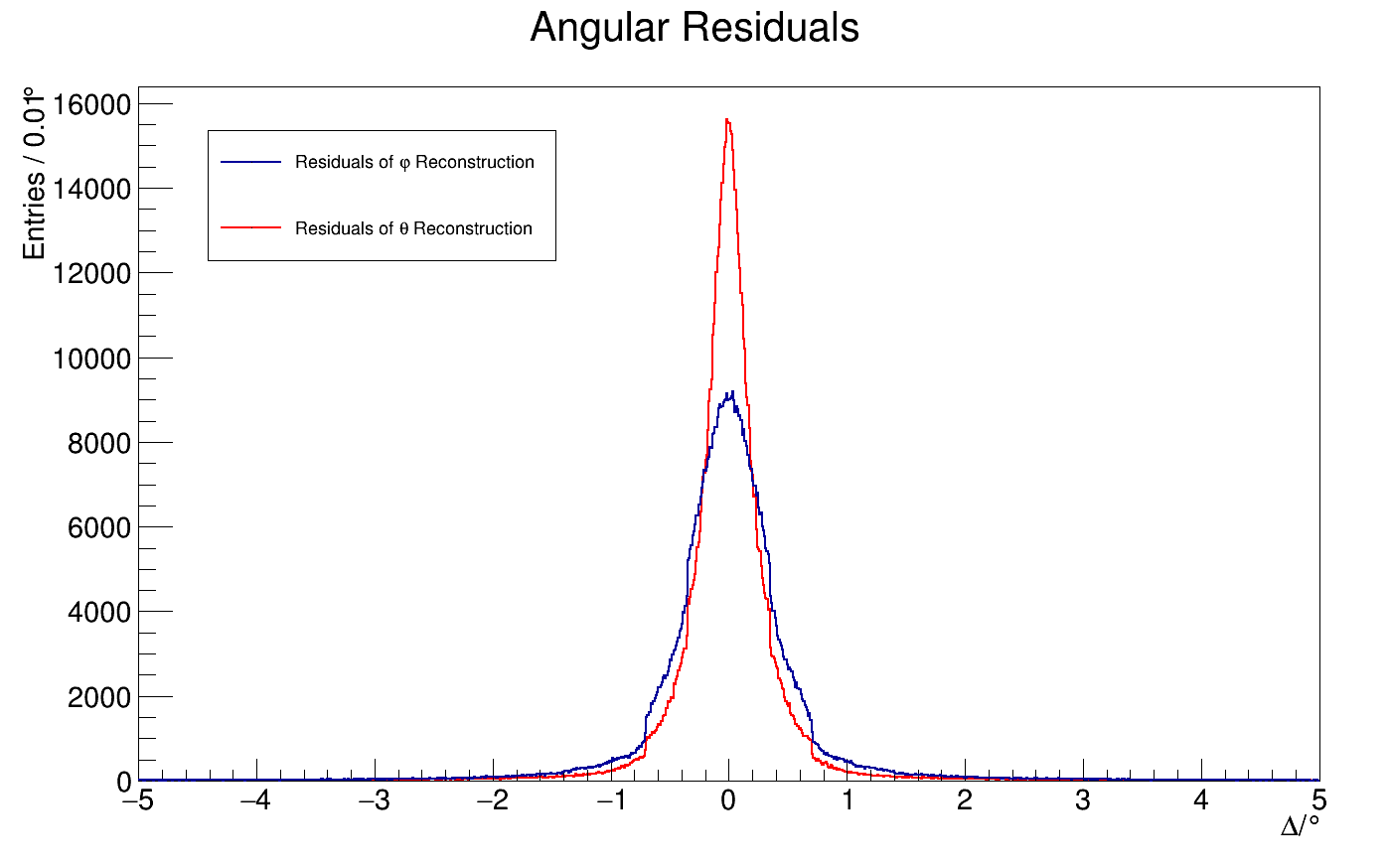}
  \caption{Angular residuals $\Delta$ of the reconstructed tracks, defined as difference between reconstructed and true angles, for \cphi (blue) and \ctheta (red) of the cases. 
  In 92\% (\cphi) and 93\% (\ctheta), respectively, the reconstructed values show a deviation of less than $1\degree$ from the true track. 
  The sharp edges edges at $~\pm0.35\degree$ and $~\pm0.7\degree$ in both distributions are caused by the discrete Hough space. 
  }
  \label{fig:angresiduals}
\end{figure}

Figure \ref{fig:angresiduals} shows the performance of the track reconstruction for the azimuthal and the polar angle: 92\% and 93\% of all reconstructed tracks are within $1\degree$ of the true track in \cphi and \ctheta, respectively. 

\begin{figure}[!ht]
  \centering
  \includegraphics[width=0.7\textwidth]{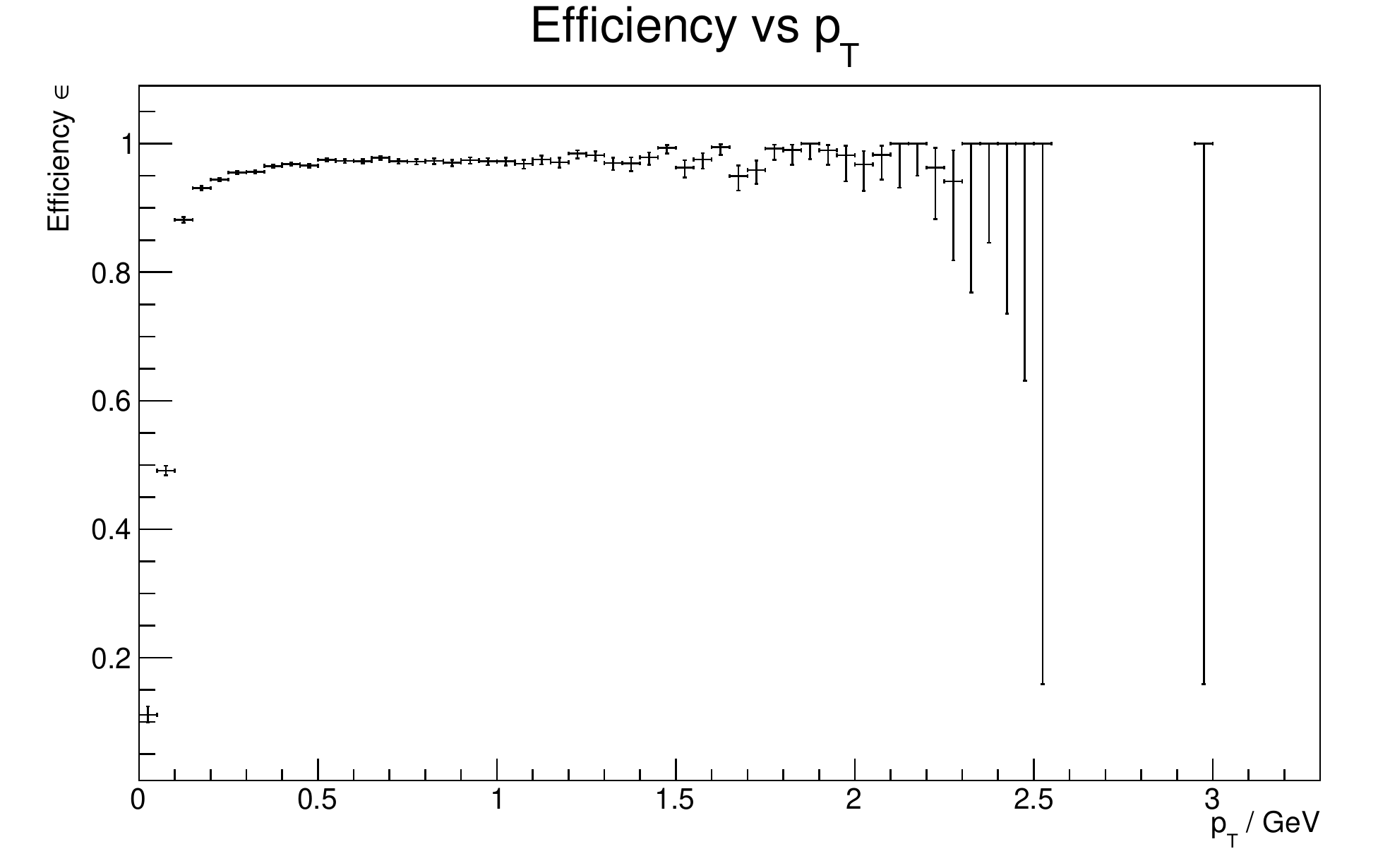}
  \caption{Track reconstruction efficiency as a function of the transverse momentum \pT. 
  The sharp drop in the low-\pT\, region is caused by the fact that particles with \pT\,< 80\,MeV do not reach the third layer of the SVD and thus do not produce the required minimum number of three SVD hits. 
  The large error bars for \pT\, > 1.5\,GeV are due to limited occurrence of particles with such a transverse momentum in the simulation. 
  The average track reconstruction efficiency is 96\% in the complete \pT-range shown.   
  }
  \label{fig:pTEfficiency}  
\end{figure}

In Figure \ref{fig:pTEfficiency}, the track reconstruction efficiency as a function of the transverse momentum, \pT, is shown: 
The overall track reconstruction efficiency is 96\% over the complete momentum range expected for the decay products of the \YfourS resonance. 
The reduced efficiency for low-\pT\, tracks is primarily due to the fact that they only traverse one or two SVD layers and the DATCON algorithm requires at least three hits in different layers to identify a track to reduce combinatorics. 

\begin{figure}[!ht]
  \centering
  \includegraphics[width=\textwidth]{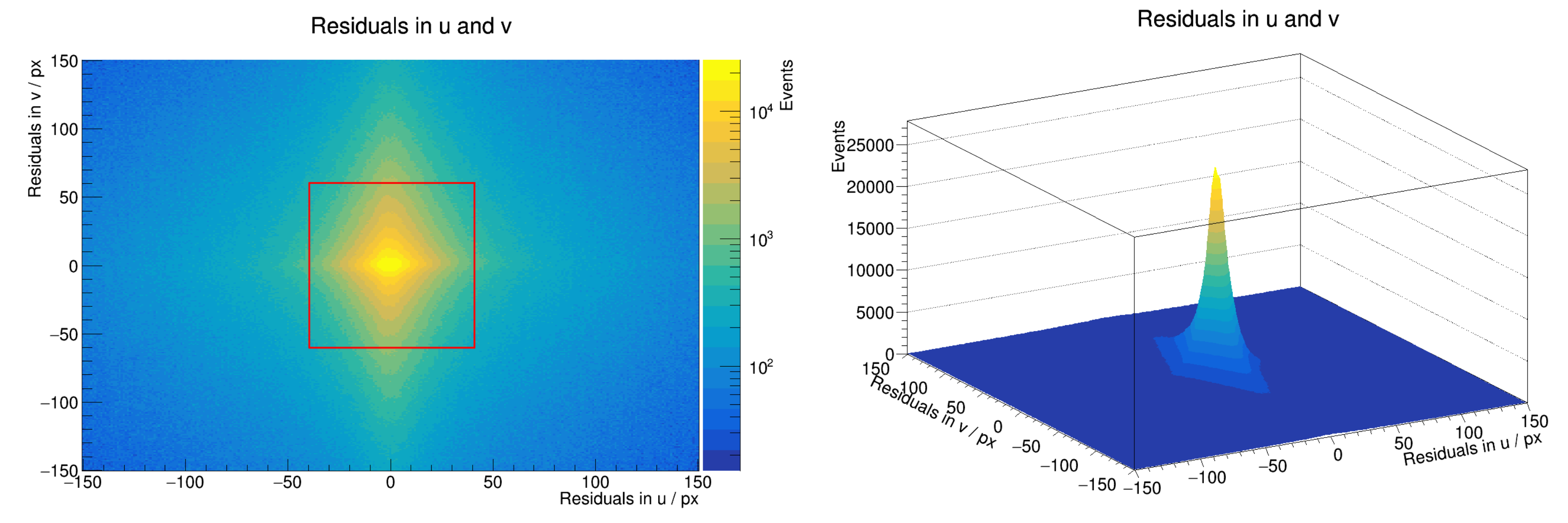}
  \caption{Residuals of the extrapolation to the PXD (measured in pixels) in local sensor coordinates $u$ and $v$. 
    The residual is defined as the difference between the local coordinates of the extrapolated hit and the local coordinates of the true simulated Monte-Carlo hit. 
    The left side shows the residuals on a logarithmic scale and the ROI of 80$\times$120 pixels around the MPH as a red box. 
    The right side shows a 3D illustration of the residuals. 
    With ROIs of size 80$\times$120 pixels, the ROI finding efficiency is 94\% with a data reduction factor of 15. 
    To obtain a higher ROI finding efficiency, it is possible to further increase the ROI size while still reducing the data by a factor of about 10. 
  }
  \label{fig:pixelresiduals}
\end{figure}

\begin{figure}[!ht]
  \centering
  \includegraphics[width=0.8\textwidth]{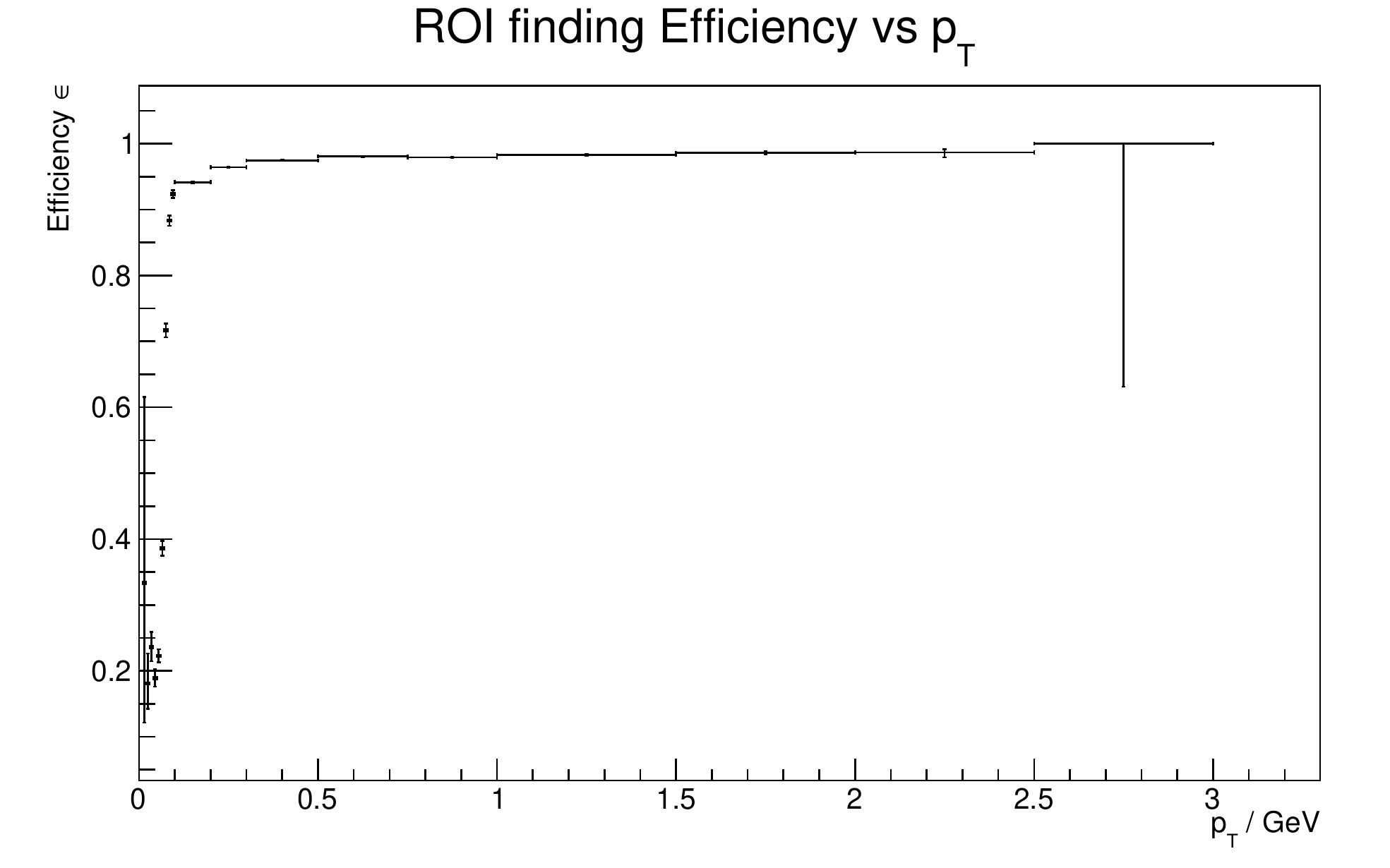}
  \caption{Efficiency of the ROI finding as a function of track \pT. 
  Particles with a transverse momentum below 80\,MeV cannot be found during tracking, and thus also no extrapolation is possible for these tracks. 
  For tracks with \pT\,> 100\,MeV, the ROI finding efficiency is above 90\%, the average over the complete \pT-range is larger than 94\%, with the data reduction factor being above 10. }
  \label{fig:ROIEfficiencyPT}
\end{figure}

Figure \ref{fig:pixelresiduals} shows the residuals of the extrapolated hits, defined as the difference between the local position of the extrapolated hit and the local position of the true MC hit, measured in pixels in the local coordinates $u$ and $v$. 
Over 90\% of the extrapolated hits are very close to the true hit position within  $\Delta R = \sqrt{u^2 + v^2}$ of 50 pixels, and 94\% of all MPHs in the PXD are located such that the ROIs of size $80 \times1 120$ pixels ($u \times v$) calculated around the MPHs include the corresponding true PXD hits. 
As can be seen in Figure \ref{fig:pixelresiduals}, the residual distribution is slightly wider in $v$-direction than in $u$-direction, which is why the ROI size is chosen to be larger in $v$-direction. 
Finally, Figure \ref{fig:ROIEfficiencyPT} shows the ROI-finding efficiency as a function of track \pT. 
The average ROI-finding efficiency over the whole \pT-range is larger than 94\%, and for tracks with \pT\,>100\,MeV the algorithm is nearly 100\% efficient. 
The median data reduction factor is about 15, leaving room to further optimise the ROI size and other aspects of the algorithm. 

\section{Conclusions and Outlook}
\label{sec:conclusions}
In this article the performance of the DATCON system was presented: 
DATCON performs an FPGA-based online track reconstruction to define ROIs on the Belle II Pixel Detector using reconstructed hits from the Silicon Vertex Detector. 
This allows one to reduce the amount of data needed to be stored offline. 
The currently achieved median data reduction factor is 15. 
The track reconstruction efficiency that can be achieved with \YfourS $\rightarrow\,B\bar{B}$ events is about 96\% over the full \pT\, range, as determined by simulation. 
The efficiency of finding a true hit within a DATCON ROI is 94\%. 
Future improvements on the performance could be obtained by tuning the ROI size using the \pT\, of the reconstructed tracks or by improving the clustering of the Hough space.

\section{Acknowledgements}
This work is supported by the German Federal Ministry of Education and Research (BMBF).

\bibliography{ExampleOfTex}

\end{document}